\documentclass[preprint2]{aastex}
\newcommand{\uv}{\mbox{$U\!-\!V$}}     % U-V
\newcommand{\vz}{\mbox{$V\!-\!z'$}}    % V-z'
\newcommand{\zk}{\mbox{$z'\!-\!K$}}    % z'-K
\newcommand{\reffig}[1]{Figure~\ref{fig:#1}}
\newcommand{\reftbl}[1]{Table~\ref{tbl:#1}}

\begin{document}

\title{%
The Subaru/\textit{XMM-Newton} Deep Survey (SXDS) -- VII.\\
Clustering Segregation with Ultraviolet and Optical Luminosities\\
of Lyman-Break Galaxies at $z\sim 3$
\altaffilmark{1}
}

\author{%
Makiko Yoshida~\altaffilmark{2},
Kazuhiro Shimasaku~\altaffilmark{2,3},
Masami Ouchi~\altaffilmark{4,5},
Kazuhiro Sekiguchi~\altaffilmark{6},
Hisanori Furusawa~\altaffilmark{7},
and
Sadanori Okamura~\altaffilmark{2,3}
}

\email{myoshida@astron.s.u-tokyo.ac.jp}

\altaffiltext{1}{%
Based on data collected at the Subaru Telescope,
which is operated by the National Astronomical Observatory of Japan.}
\altaffiltext{2}{%
Department of Astronomy, Graduate School of Science,
The University of Tokyo, Tokyo 113-0033, Japan}
\altaffiltext{3}{%
Research Center for the Early Universe, Graduate School of Science,
The University of Tokyo, Tokyo 113-0033, Japan}
\altaffiltext{4}{%
Observatories of the Carnegie Institution of Washington,
813 Santa Barbara Street, Pasadena, CA 91101, USA}
\altaffiltext{5}{%
Carnegie Fellow}
\altaffiltext{6}{%
Optical and Infrared Astronomy Division,
National Astronomical Observatory of Japan,
2-21-1, Osawa, Mitaka, Tokyo 181-8588, Japan}
\altaffiltext{7}{%
Subaru Telescope, National Astronomical Observatory of Japan, 
650 N. A'ohoku Place, Hilo, HI 96720, USA}

%---------------------------------------------------------------------
\begin{abstract}

We investigate clustering properties of Lyman-break galaxies (LBGs)
at $z\sim 3$ based on deep multi-waveband imaging data
from optical to near-infrared wavelengths
in the Subaru/\textit{XMM-Newton} Deep Field.
The LBGs are selected by \uv and \vz\ colors in one contiguous area
of 561 arcmin$^2$ down to $z'=25.5$.
We study the dependence of the clustering strength on rest-frame
UV and optical magnitudes, which can be indicators of star formation
rate and stellar mass, respectively.
The correlation length is found to be a strong function of
both UV and optical magnitudes with brighter galaxies being
more clustered than faint ones in both cases.
Furthermore, the correlation length is dependent on
a combination of UV and optical magnitudes in the sense that
galaxies bright in optical magnitude have large correlation lengths
irrespective of UV magnitude,
while galaxies faint in optical magnitude have correlation lengths
decreasing with decreasing UV brightness.
These results suggest that galaxies with large stellar masses
always belong to massive halos in which they can have various
star formation
rates, while galaxies with small stellar masses reside in less massive
halos only if they have low star formation rates.
There appears to be an upper limit to the stellar mass
and the star formation
rate which is determined by the mass of hosting dark halos.

\end{abstract}

%---------------------------------------------------------------------

%---------------------------------------------------------------------
%---------------------------------------------------------------------
\section{INTRODUCTION} \label{sec:intro}

Galaxy evolution in dark halos is a central issue of cosmology
in the Cold Dark Matter (CDM) universe.
A critical key to the problem is to reveal
what kind of dark halos host what kind of galaxies,
specifically, the relationship between the mass of dark halos
and fundamental quantities of galaxies
such as star formation rate and stellar mass.
The clustering strength of galaxies can be used to infer
the mass of hosting dark halos,
since in the standard theoretical framework of galaxy evolution,
the large-scale distribution of galaxies
is determined by the distribution of underlying dark halos,
whose spatial clustering depends on their mass,
with more massive halos having stronger spatial clustering
\citep{mo1996}.

A number of studies have examined clustering properties of
various high-redshift galaxies
\citep{giavalisco1998,giavalisco2001,foucaud2003,daddi2003,
ouchi2004a,ouchi2005,adelberger2005a,lee2006,kashikawa2006,
hildebrandt2007,quadri2007,ichikawa2007}.
Measuring clustering strength requires a large galaxy sample
from a wide sky coverage.
One considerable case in which the relationship
between dark halos and galaxies has been successfully explored
by means of clustering analysis
is that of Lyman-break galaxies (LBGs).
LBGs, which are selected by their continuum features in rest-frame UV
spectra (the Lyman-break, the Ly$\alpha$ forest, and otherwise nearly
flat continuum longward of the Ly$\alpha$) redshifted into optical
bandpasses \citep{guhathakurta1990, steidel1996},
are normal, young star-forming galaxies with modest dust extinction.
Requiring only optical imaging in a few bands,
the LBG selection method
has yielded the largest and well-controlled samples
of galaxies in the young universe at $2\lesssim z\lesssim 6$
\citep[e.g.,][]{steidel1999,steidel2003,ouchi2004b,giavalisco2004,
dickinson2004,sawicki2006,yoshida2006,iwata2007,bouwens2007}.
LBGs are as numerous as the present-day galaxies,
suggesting that they play a significant role
in the early stage of galaxy evolution.
For a review of LBGs including the history of their discovery,
see \citet{giavalisco2002}.

An important piece of evidence which links
the mass of hosting dark halos to physical properties of LBGs
has been first discovered for the UV luminosity;
the clustering strength of LBGs increases strongly with UV luminosity
\citep[e.g.,][]{giavalisco2001,ouchi2004a,adelberger2005a,
lee2006,hildebrandt2007}.
Since UV luminosity is sensitive to star formation rate,
this implies that the star formation activity of LBGs
is somehow controlled by the mass of dark halos which host them.
To make a next step forward in our understanding of
how the evolution of galaxies is related to the mass of dark halos,
it would be necessary to examine the clustering strength
as a function of other properties of galaxies
like stellar mass, age, and dust extinction.
In particular, stellar mass is a robust quantity
in terms of the star formation history of galaxies
compared to star formation rate, which can vary with time.
This kind of analysis is, however, not easy because
it requires deep and wide-field observational data
covering wavelengths simultaneously from optical to near-infrared.
Only the dependence of clustering strength on rest-frame
optical luminosity has been examined by \citet{adelberger2005b}
for 300 galaxies at $z\sim 2$ (BX objects)
selected by UV wavelength in a similar manner to LBGs.

Recently, clustering properties of near-infrared selected galaxies
at high redshift
have been examined against various quantities of galaxies
\citep{quadri2007,ichikawa2007}.
The selection based on a near-infrared magnitude allows us to
compile a nearly stellar mass-selected sample.
However, the studies at high redshift on the basis of near-infrared
selected galaxies have so far been inevitably limited to shallow
\citep[$K\lesssim 23.0$;][]{quadri2007}
or small-field \citep[$\lesssim 25$ arcmin$^2$;][]{ichikawa2007}
samples.
This is due to the lower sensitivity and smaller area coverage of
near-infrared cameras compared to optical ones.

This paper studies the dependence of
clustering strength on both UV (apparent $z'$ band) and
optical (apparent $K$ band) luminosities
for LBGs at $z\sim 3$
to investigate how the stellar mass and the star formation rate
of galaxies depend on the mass of dark halos which host them
at high redshift.
The redshift $z\sim 3$ is the best target for such studies,
since it is the highest redshift at which ground-based
near-infrared imaging can access rest-frame optical wavelengths.
To construct a large sample of LBGs at $z\sim 3$,
we performed $U$-band imaging observation with Subaru/Suprime-Cam
in the Subaru/\textit{XMM-Newton} Deep Field (SXDF),
which has a set of very deep optical multi-waveband imaging data
from the Subaru/\textit{XMM-Newton} Deep Survey
\citep[SXDS;][]{sekiguchi},
and deep $J$ and $K$ data
taken with the wide-field near-infrared camera
WFCAM on UKIRT by the UKIDSS Ultra Deep Survey
\citep[UDS;][]{lawrence2007,warren2007}.

The outline of this paper is as follows:
In \S\ref{sec:data}, a brief account of the observations
and the data is presented.
A large sample of LBGs at $z\sim 3$ is constructed in \S\ref{sec:lbg}.
Simulations to assess the redshift distribution function
and the contamination by interlopers of the sample
are also described.
Clustering analysis is made in \S\ref{sec:acf}.
A summary and conclusions are given in \S\ref{sec:sum}.

Throughout this paper, the photometric system is based on AB magnitude
\citep{oke1983}.
The cosmology adopted is a flat universe with
$\mathrm{\Omega}_m = 0.3$, $\mathrm{\Omega}_\mathrm{\Lambda} = 0.7$,
%{\bfseries
$\sigma_8 = 0.9$, baryonic density $\mathrm{\Omega}_b = 0.04$,
and a Hubble constant of $H_0 = 100\ h$ km s$^{-1}$ Mpc$^{-1}$
with $h=0.7$.
The correlation length is expressed in units of $h^{-1}$ Mpc
to facilitate comparison with previous results.

%---------------------------------------------------------------------
%---------------------------------------------------------------------
\section{DATA} \label{sec:data}

We carried out $U$-band imaging observations
in the southern part of the Subaru/\textit{XMM-Newton} Deep Field
(SXDF) [$02^h18^m00^s$, $-05^\circ00'00''$ (J2000)]
with Subaru/Suprime-Cam in 2002.
The survey field was covered with one single pointing of the
Suprime-Cam,
i.e., $34' \times 27'$, with a pixel scale of $0.''202$ pixel$^{-1}$.
The stacked image has a seeing size, PSF FWHM, of $1.35''$.
The total exposure time was $3.7$ hours
and the $1\sigma$ surface brightness fluctuation within a $1''$
diameter aperture is $29.07$ mag.
The $5\sigma$ limiting magnitude within a $2''$ diameter aperture
is $26.41$ mag.
The SXDF has a set of very wide and deep
multi-waveband optical imaging data
from the Subaru/\textit{XMM-Newton} deep survey project
\citep[SXDS;][]{furusawa2007}
taken with Subaru/Suprime-Cam in five standard broad-band filters,
$B$, $V$, $R$, $i'$, and $z'$.
We combine these data with the $U$-band data
to select LBGs at $z\sim 3$.
All of the images were aligned and smoothed with Gaussian kernels
so that all have the same PSF size as that of the $U$-band image.
The surface brightness limits ($1\sigma$ fluctuation
within a $1''$ diameter
aperture) are 30.94, 30.38, 30.17, 29.90, and 28.87 mag
in $B$, $V$, $R$, $i'$, and $z'$, respectively.
The $5\sigma$ limiting magnitudes within a $2''$ diameter aperture
are 27.88, 27.30, 27.09, 26.77, and 25.76 mag
in $B$, $V$, $R$, $i'$, and $z'$, respectively.

Part of the SXDF was imaged in the $J$ and $K$ bands
with UKIRT/WFCAM by the UKIDSS Ultra Deep Survey
\citep[UDS;][]{lawrence2007,warren2007}.
In this study we make use of these data (DR1)
to examine near-infrared properties of LBGs at $z\sim 3$.
The $J$-band and $K$-band images were also aligned
and smoothed to be matched with the optical images.
The surface brightness limits ($1\sigma$ fluctuation
within a $1''$ diameter
aperture) are 26.78 and 26.56 mag in $J$ and $K$, respectively.
The $5\sigma$ limiting magnitudes within a $2''$ diameter aperture
are 23.67 and 23.47 mag in $J$ and $K$, respectively.

Object detection and photometry were performed using SExtractor
version 2.3 \citep{bertin1996}.
We detected objects in the $z'$-band image,
and for each detected object
photometry was made in all the images at exactly the same position
by running SExtractor in ``double-image mode''.
We adopt MAG\_AUTO in SExtractor for total magnitudes,
and use magnitudes within a $2''$ diameter aperture to derive colors
\footnote{We also measure magnitudes within a larger aperture of
$3''$ diameter,
as the PSF FWHM of the final images is somewhat large.
However, since over 90 \% of the LBG candidates selected with
$2''$-aperture magnitudes overlap with those
selected with $3''$-aperture magnitudes,
we adopt $2''$-aperture magnitudes
in order to obtain colors of faint objects with better S/N.
In addition, we compare aperture magnitudes with isophotal magnitudes
and find that the isophotal areas of objects satisfying the magnitude
threshold for LBG selection ($23.0<z'\le 25.5$)
are mostly larger than the $2''$-diameter aperture.
This means that $2''$-aperture magnitudes are less
noisy than isophotal magnitudes for most objects.}
with an aperture correction of $-0.2$ mag for $U$ magnitudes
since the PSF shape in the $U$ band image was different from
those in the others.
The value $-0.2$ was determined so that the difference between
the aperture magnitudes and the total magnitudes of the $U$ band
became equal to those of the other bands for LBG candidates.
The magnitudes of objects were corrected for a small amount of
foreground Galactic extinction using the dust map of
\citet{schlegel1998}.
The reddening is $E(B-V)= 0.023$, corresponding to extinctions of
$A_U = 0.10$, $A_B = 0.09$, $A_V = 0.07$, $A_R = 0.06$, $A_{i'} = 0.05$,
$A_{z'} = 0.03$, $A_J = 0.02$, and $A_K = 0.01$.

Clustering analysis requires high uniformity in sensitivity
over the whole area, since fluctuations of sensitivity can produce
spurious clustering signals
and bias measurements of clustering strength.
We examined the sensitivity variation over the area of the images
of respective bands by dividing them into small meshes
and estimating the sky noise
in each of the meshes.
Based on these sky-noise maps,
we carefully defined a high-quality region
in which the sensitivity is good and uniform,
trimming the edges of the images where sky noise was systematically
larger due to dithering observation.
The effective area with a complete coverage in all of the six
optical bands amounts to 740 arcmin$^2$,
and the overlapping area of the optical images and
the near-infrared images
is 561 arcmin$^2$ after the low-quality regions are discarded.

Spectroscopic follow-up observations have been carried out
for many objects in the SXDF with Subaru/FOCAS and VLT/VIMOS
\citep{akiyama,simpson,saito2007}.
The number of spectroscopically observed objects
which are located in the region used in this study
and have magnitudes in the range of $23.0<z'\le 25.5$,
within which we select LBGs,
is 63.

%---------------------------------------------------------------------
%---------------------------------------------------------------------
\section{LYMAN-BREAK GALAXY SAMPLE AT $z\sim 3$}
\label{sec:lbg}

\subsection{Selection of Lyman-break Galaxies}
\label{subsec:lbg-select}

\begin{figure}[t]
  \begin{center}
    \includegraphics[scale=0.4]{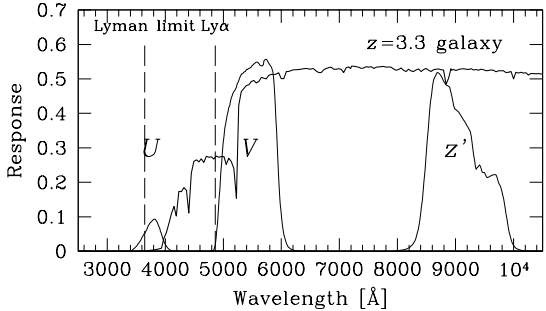}
  \end{center}
  \caption{%
The $U$, $V$, and $z'$ bandpasses
overplotted on the spectrum of a generic $z=3$ galaxy
(thick line), illustrating the utility of color
selection technique using these three bandpasses for
locating $z\sim 3$ galaxies.
\label{fig:sed}}
\end{figure}

\begin{figure}[p]
  \begin{center}
    \includegraphics[height=7cm]{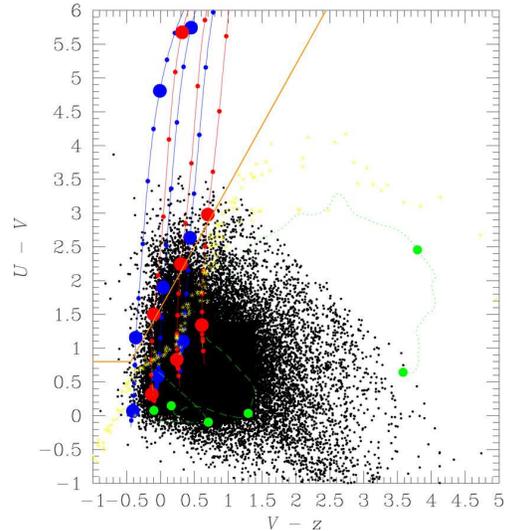}
  \end{center}
  \caption{%
\uv\ vs \vz\ diagram for the detected objects with $23.0 < z' < 25.5$.
When the $U$ and/or $V$ magnitude of an object is fainter
than the $1\sigma$ magnitude of the band, the $1\sigma$
magnitude is assigned to the object.
The predicted colors of model galaxies and stars are overplotted.
The blue and red solid lines indicate the tracks for model spectra of
young star-forming galaxies of age 0.1 Gyr and 1.5 Gyr, respectively,
with reddening of $E(B-V)=0$, $0.15$,
and $0.3$ (from left to right).
The redshift range is from $z=2$ to higher redshifts,
and the circles on the track mark the redshift interval of 0.1
with the enlarged circles corresponding to $z=2.5$, $3.0$, and $3.5$.
The green dotted, dashed, and dot-dashed lines delineate the tracks for
model spectra of local elliptical, spiral, and irregular galaxies,
respectively, redshifted from $z=0$ to $2$ without evolution.
The circles on each track mark $z=0$, $1$, and $2$.
The yellow asterisks represent the colors of 175 Galactic stars
given by \citet{gunn1983}.
The thick orange line indicates the boundary which we adopt for
the selection of $z\sim 3$ LBGs.
\label{fig:color}}
\end{figure}

\begin{figure}[t]
  \begin{center}
    \includegraphics[scale=0.4]{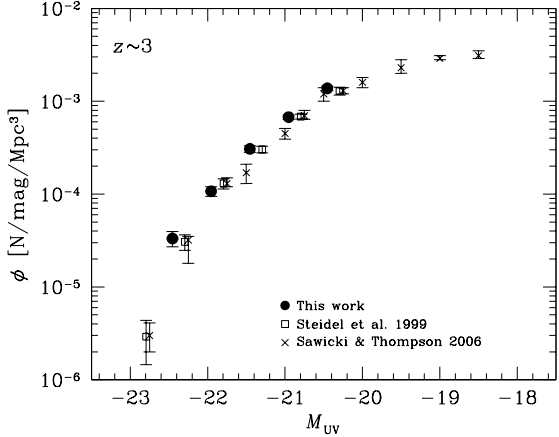}
  \end{center}
  \caption{%
UV luminosity function for LBGs at $z\sim 3$.
Our data are shown by the filled circles.
The open squares and crosses are from \citet{steidel1999}
and \citet{sawicki2006}, respectively.
\label{fig:lf}}
\end{figure}
  
\begin{figure}[t]
  \begin{center}
    \includegraphics[scale=0.4]{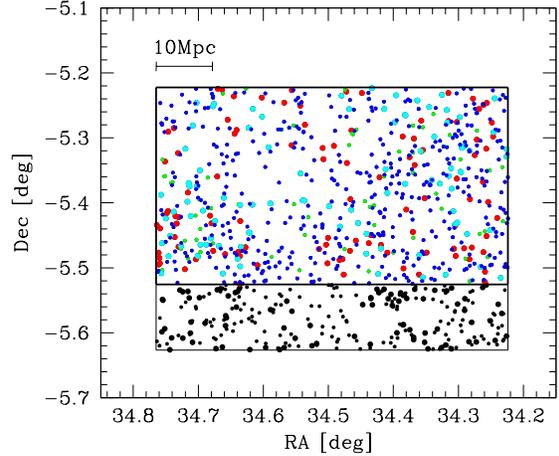}
  \end{center}
  \caption{%
Sky distribution of the 962 LBGs in our sample.
The region outlined by the thick line indicates the area observed
in the $J$ and $K$ bands,
while the rest is the area observed only in optical bands,
after removal of low-quality regions.
The large and small circles represent LBGs with $23.0<z'\le 24.5$
and $24.5<z'\le 25.5$, respectively.
LBGs having $K$ magnitudes are further distinguished
by $z'$ and $K$ magnitudes with different colors,
where red, green, cyan, and blue represent LBGs with
$23.0<z'\le 24.5$ and $K\le 23.46$, $24.5<z'\le 25.5$ and $K\le 23.46$,
$23.0<z'\le 24.5$ and $K>23.46$, $24.5<z'\le 25.5$ and $K>23.46$,
respectively.
The projected comoving scale of 10 Mpc at $z=3.3$
is shown at the top left.
North is up and east is to the left in this image.
\label{fig:space}}
\end{figure}
  
We find that a combination of $U$, $V$, and $z'$ bands works best
to select LBGs at $z\sim 3$ among our bandpasses set (\reffig{sed}).
Note that the $U$-band filter of the Suprime-Cam is significantly
redder than standard $U$-band filters.
Consequently, the mean redshift of LBGs selected is
higher than those of traditional $U$-drop LBG samples
(see \S\ref{subsec:lbg-comp}).
In \reffig{color}, we show the distribution of detected objects
with $23.0 < z' \le 25.5$ in the \uv vs \vz\ diagram,
as well as predicted positions of high-redshift galaxies
and foreground objects (lower-redshift galaxies and Galactic stars).
When the magnitude of an object in $U$ and/or $V$ is fainter
than the $1\sigma$ magnitude,
the $1\sigma$ magnitude is assigned.

We set the selection criteria for LBGs at $z\sim 3$ as:
\begin{mathletters}
  \begin{eqnarray}
      && 23.0 < z' \le 25.5, \\
%      && U-V \ge 0.8,\ \ \ V-z' \le 2.7,\ \ \ U-V \ge 1.8(V-z') + 1.6.
      && U-V \ge 0.8,\ \ \ V-z' \le 2.7, \nonumber \\
      && \hspace{0.5cm} U-V \ge 1.8(V-z') + 1.6.
  \end{eqnarray}
\end{mathletters}
The boundaries on the \uv\ vs \vz\ diagram
defined by these color criteria
are outlined with the thick orange line in \reffig{color}.
The number of LBG candidates selected is 962 in total, among which
$708$ are found in the region observed in the $J$ and $K$ bands.

Among the 63 spectroscopic objects, 4 objects
satisfy the selection criteria and all of them are identified to be
at $z>2.9$.
On the other hand,
there are two additional objects which are found
in the redshift range of $2.9<z<3.7$ but do not pass the
criteria.
These two are
fainter in $U$ than the $1\sigma$ magnitude and are not red enough
in $U_{1\sigma}-V$ color.
The possibility of missing targeted galaxies for this reason
is taken into account in a simulation for estimating the completeness
of our sample described in \S\ref{subsec:lbg-comp}.
The missed objects have $z'$ magnitudes of $z'=24.93$ and $25.13$.
In fact, the simulation shows that
a small fraction of $z\sim$ 40\% of galaxies
at $2.9<z<3.7$ with these magnitudes is selected by the criteria.

The luminosity function for the LBG sample at $z\sim 3$ is
derived in the same manner as in \citet{yoshida2006}
using the completeness and the contamination fraction
obtained in the following subsections
(\S\ref{subsec:lbg-comp} and \S\ref{subsec:lbg-contam}).
\reffig{lf} shows the luminosity function
in comparison with those in the literature
\citep{steidel1999,sawicki2006}.
Note that the absolute UV luminosities of our galaxies
are based on apparent
$z'$ magnitudes, while others are based on apparent
$R$ magnitudes.
The agreement of our luminosity function with
those in previous studies is very good,
particularly when one considers different sample selections.
\reffig{space} shows the sky distribution of the LBG sample.

\subsection{Redshift Distribution Function} \label{subsec:lbg-comp}

\begin{figure}[t]
  \begin{center}
    \includegraphics[scale=0.4]{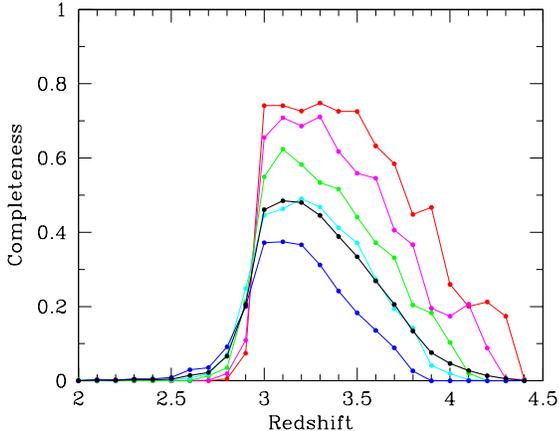}
  \end{center}
  \caption{%
Redshift distribution functions, $N(z)$, of the LBGs
with different apparent magnitudes
estimated from Monte Carlo simulations.
The red, magenta, green, cyan, and blue lines denote the $N(z)$
for $z'=23.25$, 23.75, 24.25, 24.75, 25.25, respectively.
The thick black line indicates the magnitude-weighted
redshift distribution function.
\label{fig:complete}}
\end{figure}
  
The redshift distribution function of the LBG sample is estimated
as a function of magnitude through a Monte Carlo simulation.
In the simulation, we generate artificial LBGs over an apparent
magnitude range of $23.0\le m_{z'}\le 25.5$
with an interval of $\Delta m=0.5$, and over a redshift range of
$2.0\le z\le 4.5$ with an interval of $\Delta z = 0.1$.
Model spectra of the artificial LBGs are constructed
using the stellar population synthesis code
developed by \citet{kodama1997}.
As model parameters, an age of 0.1 Gyr, a Salpeter initial mass
function, and a star-formation timescale of 5 Gyr are adopted,
and five values of reddening, $E(B-V) = 0.0$, $0.1$, $0.2$, $0.3$,
and $0.4$,
are applied using the
dust extinction formula for starburst galaxies by
\citet{calzetti2000}.
These values reproduce the average rest-frame ultraviolet-optical
spectral energy distribution
of LBGs observed at $z\sim 3$ \citep{papovich2001}.
The absorption due to the intergalactic medium is applied
following the prescription by \citet{madau1995}.
The colors are calculated by convolving thus
constructed model spectra with the response functions of the
Suprime-Cam filters.
We assume that the surface brightness distribution of LBGs is Gaussian
and assign apparent sizes to the artificial LBGs so that their size
distribution measured by SExtractor
matches that of the observed LBG candidates.
The artificial LBGs are then distributed randomly on the original
images after adding Poisson noise according to their magnitudes, and
object detection and photometry are performed in the same manner
as done for real objects.
A sequence of these processes is repeated 100 times to obtain
statistically accurate values of completeness.
In the simulation, the completeness for a given apparent magnitude,
redshift, and $E(B-V)$ value
can be defined as the ratio in number of the simulated LBGs
which are detected and also satisfy the selection criteria, to all
the simulated objects with the given magnitude, redshift,
and $E(B-V)$ value.
We calculate the completeness of the LBG sample
by taking a weighted average of the completeness
for each of the five $E(B-V)$ values.
The weight is taken using the $E(B-V)$ distribution function of
$z\sim 4$ LBGs derived by
\citet{ouchi2004b} (open histogram in the bottom panel of their Fig.20),
which has been corrected for
incompleteness due to selection biases.
The resulting completeness, $p(m,z)$, is shown in \reffig{complete}.

The magnitude-weighted redshift distribution function
of our LBG sample is derived from $p(m,z)$
by averaging the magnitude-dependent completeness
weighted by the number of LBGs in each magnitude bin
(a thick black line in \reffig{complete}).
The average redshift, $\bar{z}$, and its standard deviation, $s_z$,
are calculated to be $\bar{z}=3.3$ and $s_z=0.3$.
As we mentioned in \S\ref{subsec:lbg-select},
due to the redder $U$-band filter of the Suprime-Cam,
the redshift distribution function is biased toward
higher redshifts in comparison with traditional $U$-drop LBGs,
whose redshifts are distributed around $z\sim 3.0$
\citep[e.g.,][]{steidel2003}.

The measurement of the correlation length from the angular correlation
function relies on the estimation of the redshift distribution function
(\S\ref{sec:acf}).
In order to explore to what extent the redshift distribution function
and the resultant correlation length are affected
by the model assumption,
we recalculate the redshift distribution function
using a different model spectrum with an age of 1.5 Gyr.
In fact, some of the LBGs in our sample have red \zk colors explained
by such spectra.
The derived correlation lengths based on the two different model
spectra are found to be consistent within the errors.

\subsection{Contamination by Interlopers} \label{subsec:lbg-contam}

\begin{figure}[t]
  \begin{center}
    \includegraphics[scale=0.4]{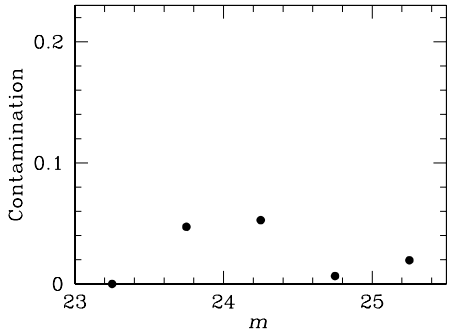}
  \end{center}
  \caption{%
Fraction of interlopers as a function of magnitude for the LBG sample.
\label{fig:contam}}
\end{figure}

We estimate the fraction of low-redshift interlopers in the LBG
sample also by a Monte Carlo simulation as follows.
The adopted boundary redshift between interlopers and LBGs is
$z_0=2.9$.
We use objects in the Hubble Deep Field North (HDFN), for which
best-fit spectra and photometric redshifts are given by
\citet{furusawa2000}, as a template of the color, magnitude, and
redshift distribution of foreground galaxies, and generate 929
artificial objects which mimic the HDFN objects.
The apparent sizes of the artificial objects are adjusted so that
the size distribution recovered from the simulation is similar to
that of the real objects in our catalogs.
We distribute the artificial objects randomly on the original images
after adding Poisson noise according to their magnitudes, and
perform object detection and photometry in the same manner as
employed for real objects.
A sequence of these processes is repeated 100 times.
In the simulation, the number of interlopers can be defined as the
number of the simulated objects with low redshifts ($z<z_0$) which are
detected and also satisfy the selection criteria for LBGs.
The number of interlopers expected in the LBG sample can then be
calculated by multiplying the raw number by a scaling factor
which corresponds to the
ratio of the area of our field (740 arcmin$^2$) to
the area of the HDFN
multiplied by the repeated times (100 $\times $ 3.92 arcmin$^2$).
\reffig{contam} shows the fraction
of interlopers for our LBG sample as a function of magnitude.
The fraction is found to be at most 6\% at any magnitude.

%---------------------------------------------------------------------
%---------------------------------------------------------------------
\section{CLUSTERING PROPERTIES} \label{sec:acf}

\subsection{Method} \label{subsec:acf-method}

We measure the angular correlation function (ACF), $\omega(\theta)$,
using the estimator proposed by \citet{landy1993}:
\begin{eqnarray}
  \omega_\mathrm{obs}(\theta) &=&
    \frac{DD(\theta)-2DR(\theta)+RR(\theta)}{RR(\theta)},
\end{eqnarray}
where $DD(\theta)$, $DR(\theta)$, and $RR(\theta)$
denote the numbers of galaxy-galaxy, galaxy-random,
and random-random pairs, respectively,
with angular separations between $\theta-\delta\theta/2$
and $\theta+\delta\theta/2$.
The distribution of random points is subject to exactly
the same geometry
as the observed area, avoiding regions where galaxies
are not detected, e.g., in the vicinity of bright stars.
We generate 100 times as many random points as the number of galaxies
in order to reduce the uncertainties.
The errors are estimated by bootstrap resampling method
\citep{ling1986}.
As the contamination is very small for our LBG sample,
we do not apply any correction for it.

The ACF can be approximated as a power law given by
\begin{eqnarray}
\omega(\theta) &=& A_\omega\theta^{-\beta}.
\end{eqnarray}
However, since the average number of galaxies in a given field
is estimated
from a sample itself and
fluctuations on the scale of the field size are not accounted for,
the measured ACF is underestimated by a constant
known as the integral constraint, IC \citep{groth1977}:
\begin{eqnarray}
\omega_\mathrm{obs}(\theta) &=& \omega_\mathrm{true}(\theta)
  - \mathrm{IC}
  \ =\ 
  A_\omega\theta^{-\beta} - \mathrm{IC}.
\label{eq:omega}
\end{eqnarray}
The value of the integral constraint is equal to the
variance of the number of galaxies in the field:
\begin{eqnarray}
\mathrm{IC} &=& \frac{1}{N_\mathrm{gal}} + \sigma_\omega^2,
\label{eq:ic}
\end{eqnarray}
where the first term is the Poisson variance and the second term
accounts for an additional variance caused by clustering.
The variance caused by clustering can be estimated by
integrating $\omega_\mathrm{true}(\theta)$
over the survey area \citep{roche1999}:
\begin{eqnarray}
\sigma_\omega^2 &=&
  \frac{1}{\Omega^2}\int\int\omega_\mathrm{true}(\theta)\,
  d\Omega_1d\Omega_2 \nonumber \\
            &=&
  \frac{\sum_iRR(\theta_i)\omega_\mathrm{true}(\theta_i)}
       {\sum_iRR(\theta_i)} \nonumber \\
            &=&
  \frac{\sum_iRR(\theta_i)A_\omega\theta^{-\beta}}
       {\sum_iRR(\theta_i)}.
\end{eqnarray}
The quantity $\sigma_\omega^2/A_\omega$ is estimated directly
from the random point catalogs for a given value $\beta$.
Then the amplitude of the ACF, $A_\omega$, can be estimated
through $\chi^2$ fitting of $\omega_\mathrm{obs}(\theta)$
using equations (\ref{eq:omega}) and (\ref{eq:ic}).
Following recent results in clustering studies of LBGs
\citep{adelberger2005a,lee2006},
we assume $\beta = 0.6$ in what follows.
The fitting is made only at $\theta>10''$ since at $\theta<10''$
there may be a significant contribution from nonlinear small scale
clustering \citep{ouchi2005,lee2006}.
The error in $A_\omega$ is estimated from the range in which
an increase in $\chi^2$ from the best-fit value is less than unity.

We derive the spatial correlation function, $\xi(r)$,
by inverting $\omega(\theta)$ using the Limber transform
\citep{peebles1980}.
If the ACF is a power law, the spatial correlation function
also has to have a power law form:
\begin{eqnarray}
  \xi(r) &=& \left(\frac{r}{r_0}\right)^{-\gamma},
\end{eqnarray}
where $r_0$ is the spatial correlation length and $\gamma = \beta+1$.
The $A_\omega$ is related to $r_0$ as
\begin{eqnarray}
  A_\omega &=& Cr_0^\gamma\int F(z)D_\theta^{1-\gamma}(z)N(z)^2g(z)
                                           \,dz \nonumber \\
           & & \hspace{1.8cm}{}\times\left[\int N(z)\,dz\right]^{-2},
\label{eq:r0}
\end{eqnarray}
where $D_\theta$ is the angular diameter distance,
$N(z)$ is the redshift distribution function,
\begin{eqnarray}
  g(z) &=& \frac{H_0}{c}\left\{(1+z)^2(1+\Omega_0z+\Omega_\Lambda
              \left[(1+z)^{-2}-1)\right]^{1/2}\right\}, \nonumber
\end{eqnarray}
and
\begin{eqnarray}
  C &=& \sqrt{\pi}\frac{\Gamma[(\gamma-1)/2]}{\Gamma(\gamma/2)}.
  \nonumber
\end{eqnarray}
The function $F(z)$ describes the evolution of $\xi(r)$ with redshift.
The evolution is often modeled as
$F(z) = \left[(1+z)/(1+\bar{z})\right]^{-(3+\epsilon)}$.
We assume constant clustering in comoving units
in the redshift range of our LBG sample;
in this case the parameter $\epsilon$ is specified by
$\epsilon = \gamma-3$.
For the redshift distribution function, we use the one obtained by
the simulation in \S\ref{subsec:lbg-comp}.

The standard CDM model predicts that the clustering of dark halos
is correlated with halo mass \citep{mo1996}.
We use the observed spatial correlation function of LBGs to infer
the mass of dark halos hosting them
on the basis of the analytic model given by \citet{sheth2001},
which is derived from a fit to large N-body simulations.
According to the model, the bias of dark halos, $b_\mathrm{DH}$,
which relates the clustering of dark halos
to that of the overall dark matter, is calculated by
\begin{eqnarray}
b_\mathrm{DH} &=& 1 + \frac{1}{\delta_c}
\left[\nu'^2+b\nu'^{2(1-c)}\right. \nonumber \\
& & \hspace{0.5cm}\left.
    {}-\frac{\nu'^{2c}/\sqrt{a}}{\nu'^{2c}+b(1-c)(1-c/2)}\right],
\label{eq:bias_dh}
\end{eqnarray}
where $\nu' = \sqrt{a}\nu$, and the constants
$a = 0.707$, $b = 0.5$, $c = 0.6$.
Here, $\nu$ is defined by
\begin{eqnarray}
\nu &\equiv& \frac{\delta_c}{\sigma(M, z)}
    \ =\ \frac{\delta_c}{D(z)\sigma(M, 0)},
\label{eq:nu}
\end{eqnarray}
where $D(z)$ is the growth factor, $\sigma(M, z)$
is the relative mass fluctuation in spheres that contain
an average mass $M$,
and $\delta_c \approx 1.69$ represents the critical amplitude
of the perturbation for collapse.
We calculate $D(z)$ following \citet{carroll1992}
and $\sigma(M,0)$ from the initial power spectrum
with a power law index
of $n=1$ using the transfer function of \citet{bardeen1986}.
Since $\nu'$ is a function of redshift and mass,
the mass of dark halos is estimated from equation (\ref{eq:bias_dh}),
once the bias of the dark halos and the redshift are given.
We assume that the observed bias of galaxies at a large scale
reflects the bias of dark halos hosting them
(i.e., $b_\mathrm{gal}\simeq b_\mathrm{DH}$),
thereby obtain an estimate of the hosting halo mass.
The bias of galaxies against dark matter at a large scale
($=8\ h^{-1}$Mpc) is measured by
\begin{eqnarray}
b_\mathrm{gal} &=& \sqrt{\frac{\xi(8\ h^{-1}\mathrm{Mpc})}
  {\xi_\mathrm{DM}(8\ h^{-1}\mathrm{Mpc})}} \nonumber \\
  &=& \sqrt{\frac{\left[(8\ h^{-1}\mathrm{Mpc})/r_0\right]^{-\gamma}}
       {\xi_\mathrm{DM}(8\ h^{-1}\mathrm{Mpc})}},
\label{eq:bias_gal}
\end{eqnarray}
where $\xi_\mathrm{DM}$ is the predicted spatial correlation function
of the overall dark matter and is computed by the nonlinear model of
\citet{peacock1996}.

\subsection{Results} \label{subsec:acf-result}

\begin{figure}[t]
  \begin{center}
    \includegraphics[scale=0.4]{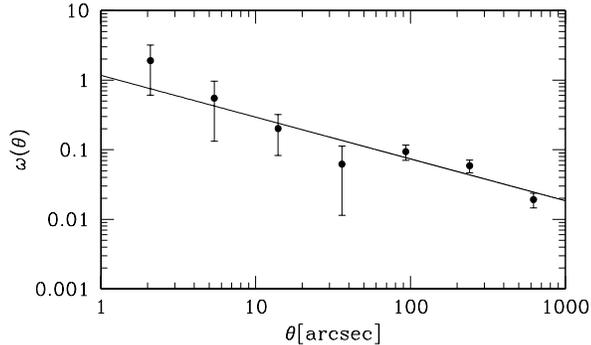}
  \end{center}
  \caption{%
Angular correlation function, $\omega(\theta)$,
for the $z\sim 3$ LBG sample.
\label{fig:acf}}
\end{figure}

\reffig{acf} shows the ACF for the full LBG sample
containing 962 galaxies
with $23.0 < z'\le 25.5$.
We calculate the spatial correlation length from the amplitude of the
ACF using equation (\ref{eq:r0}) to find
$r_0 = 5.5^{+0.8}_{-0.9}\ h^{-1}$Mpc (comoving; \reftbl{clustering}).
The average mass of dark halos hosting these LBGs is estimated to be
$M_\mathrm{DM}\approx 2\times 10^{12}M_\odot$
from equations (\ref{eq:bias_dh}), (\ref{eq:nu}), (\ref{eq:bias_gal}).
We compare the measurement of the spatial correlation length
with those from other authors for samples
with similar redshift and luminosity ranges.
\citet{hildebrandt2007} found $r_0 = 5.0^{+0.2}_{-0.2}\ h^{-1}$Mpc
for LBGs with $23.5 < R \le 25.5$, which is consistent with ours.
On the other hand, \citet{adelberger2005a} and \citet{lee2006}
found slightly smaller values of
$r_0 = 4.0^{+0.6}_{-0.6}\ h^{-1}$Mpc for LBGs with $23.5 < R \le 25.5$
and $r_0 = 4.1^{+0.1}_{-0.2}\ h^{-1}$Mpc for LBGs with $R \le 25.5$,
respectively.
This may be due in part to the fact that the redshift distribution
function of our LBG sample is biased toward higher
redshifts
as discussed in \S\ref{subsec:lbg-comp}.
The average redshift of Adelberger et al.'s (2005a) sample
is $\bar{z}=2.9$,
while ours is $\bar{z}=3.3$.
At a given magnitude limit, this means that our sample includes
brighter LBGs, which are more strongly clustered.
In addition, at a fixed luminosity,
LBGs at higher redshifts are more strongly clustered.
We also note
that our sample is $z'$-magnitude limited,
while the others are $R$-magnitude limited.
The distribution in the $z'$ vs $R$ diagram of our sample implies that
there might be more galaxies with $z'>25.5$ than those with $R>25.5$,
suggesting that our sample may systematically miss more faint galaxies
than the other samples.
This could result in our larger $r_0$.

In the following,
we measure the spatial correlation length for various subsamples
to investigate the relationship between clustering strength
and galaxy properties.

\subsubsection{Dependence of Clustering Strength on UV Luminosity}
\label{subsubsec:acf_mag_z}

\begin{figure}[t]
  \begin{center}
    \includegraphics[scale=0.4]{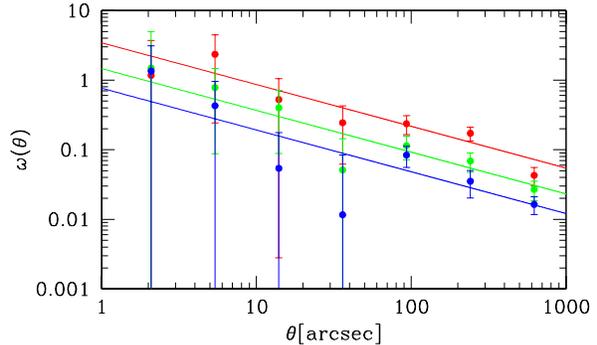}
  \end{center}
  \caption{%
Angular correlation function, $\omega(\theta)$,
for three subsamples selected by $z'$ magnitude.
The red, green, and blue symbols indicate the $\omega(\theta)$ of
LBGs with $23.0 < z' \le 24.5$, $24.0 < z' \le 25.0$,
and $24.5 < z' \le 25.5$, respectively.
\label{fig:acf_mag_z}}
\end{figure}

\begin{figure}[h]
  \begin{center}
    \includegraphics[scale=0.4]{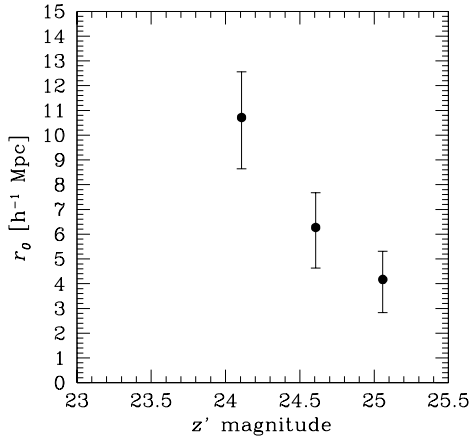}
  \end{center}
  \caption{%
Spatial correlation length, $r_0$, as a function of $z'$ magnitude.
\label{fig:r0_mag_z}}
\end{figure}

It is reported that the clustering strength of LBGs
depends on the rest-frame UV luminosity
in the sense that galaxies with higher UV luminosities
have larger spatial correlation lengths
\citep[e.g.,][]{giavalisco2001,ouchi2004a,adelberger2005a,
lee2006,hildebrandt2007}.
We examine the clustering segregation with respect to the
rest-frame UV luminosity
in our sample using subsamples selected by $z'$ magnitude
($23.0<z'\le 24.5$, $24.0<z'\le 25.0$, and $24.5<z'\le 25.5$)
from the full sample (\reftbl{clustering}).
The ACF for each subsample is shown in \reffig{acf_mag_z}.
In \reffig{r0_mag_z}, the derived correlation length is plotted
as a function of $z'$ magnitude.
The magnitude of each point in \reffig{r0_mag_z} is the median
magnitude of the corresponding subsample.
In agreement with previous studies,
we find that the correlation length increases with rest-frame
UV luminosity,
suggesting that galaxies with higher star formation rates
are hosted by more massive dark halos.

\subsubsection{Dependence of Clustering Strength on Optical Luminosity}
\label{subsubsec:acf_mag_k}

\begin{figure}[t]
  \begin{center}
    \includegraphics[scale=0.4]{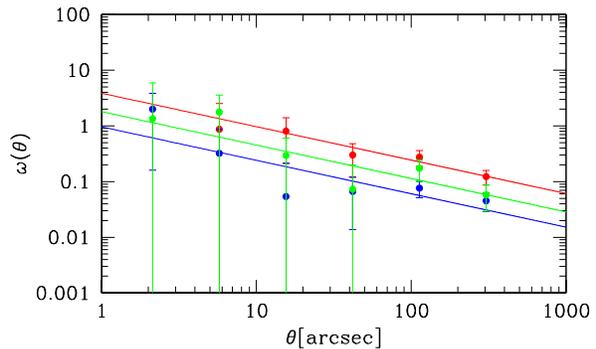}
  \end{center}
  \caption{%
Angular correlation function, $\omega(\theta)$,
for three subsamples selected by $K$ magnitude.
The red, green, and blue symbols indicate the $\omega(\theta)$ of
LBGs with $K \le 23.46$, $22.96 <K \le 23.96$  and $K > 23.46$,
respectively.
\label{fig:acf_mag_k}}
\end{figure}

\begin{figure}[h]
  \begin{center}
    \includegraphics[scale=0.4]{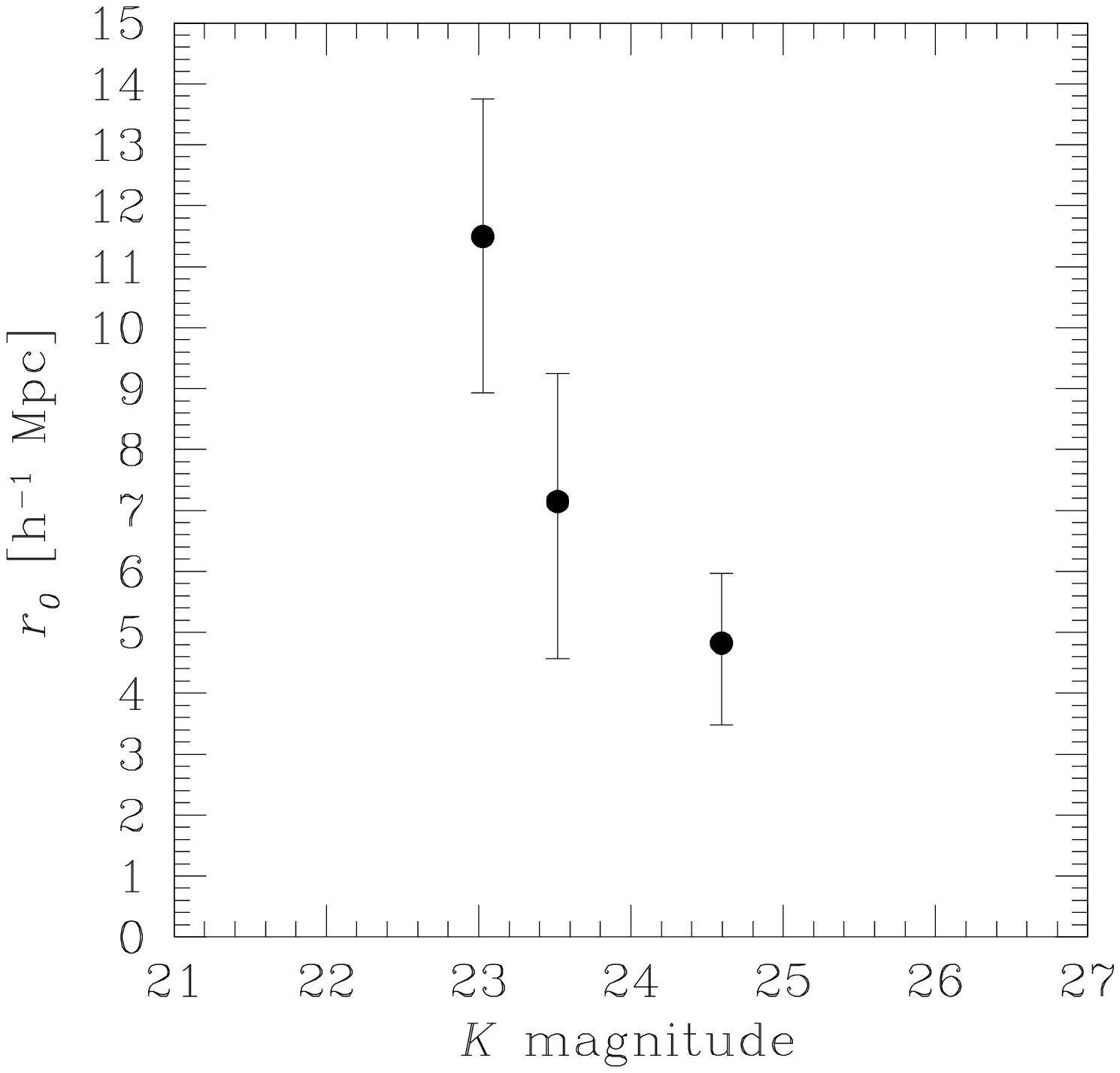}
  \end{center}
  \caption{%
Spatial correlation length, $r_0$, as a function of $K$ magnitude.
\label{fig:r0_mag_k}}
\end{figure}

\begin{figure}[h]
  \begin{center}
    \includegraphics[scale=0.4]{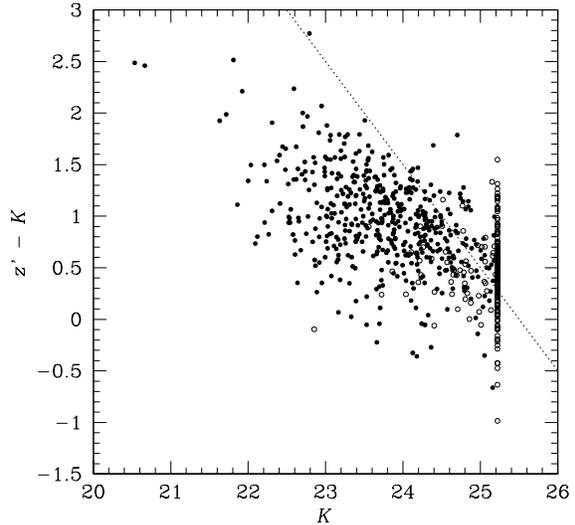}
  \end{center}
  \caption{%
\zk\ plotted against $K$ for the LBG sample.
When the $K$ magnitude of an LBG is fainter
than the $1\sigma$ magnitude,
the $1\sigma$ magnitude is assigned (open circles).
Just for a guide, the faint limit to the sample of
$z_\mathrm{total}\le 25.5$ is indicated with the dotted line.
Note that the \zk\ colors are measured with
2"-aperture magnitudes and some objects lie beyound the line.
\label{fig:k_zk}}
\end{figure}

We divide the sample by $K$ magnitude ($K\le 23.46$,
$22.96<K\le 23.96$, and $23.46<K$)
to see clustering dependence on the rest-frame optical luminosity
($\sim 5000$\AA),
which can be an indicator of stellar mass (\reftbl{clustering}).
The ACF for each subsample is shown in \reffig{acf_mag_k}.
\reffig{r0_mag_k} shows the relationship between
the derived correlation length and $K$ magnitude.
The magnitude of each point in \reffig{r0_mag_k} is the median
magnitude of the corresponding subsample.
It is found that there is a strong trend that LBGs
brighter in rest-frame optical are also more strongly clustered.

Our LBG sample shows a correlation between
$K$ magnitude and $z'-K$ color;
there is a significant deficit of bright LBGs with blue $z'-K$ colors
(\reffig{k_zk}).
Consequently, we find that correlation length increases
with $z'-K$ color from
$r_0 = 5.4^{+1.4}_{-1.7}\ h^{-1}$Mpc for $z'-K\le 0.8$ to
$r_0 = 8.8^{+1.6}_{-1.9}\ h^{-1}$Mpc for $z'-K>0.8$
(\reftbl{clustering}).
Both results on the dependence of the clustering strength imply that
galaxies with large stellar masses reside in massive dark halos.
\citet{shapley2005} suggested a relationship between stellar mass
and $R-K$ color as well as $K$ magnitude
for their sample of BX objects at $z\sim 2$,
and \citet{adelberger2005b} ascribed a similar segregation
of the clustering strength with $K$ magnitude and $R-K$ color
for the sample to an underlying correlation with stellar mass.

\subsubsection{Dependence of Clustering Strength \
on a Combination of UV and Optical Luminosity}
\label{subsubsec:acf_mag_zk}

\begin{figure}[t]
  \begin{center}
    \includegraphics[scale=0.4]{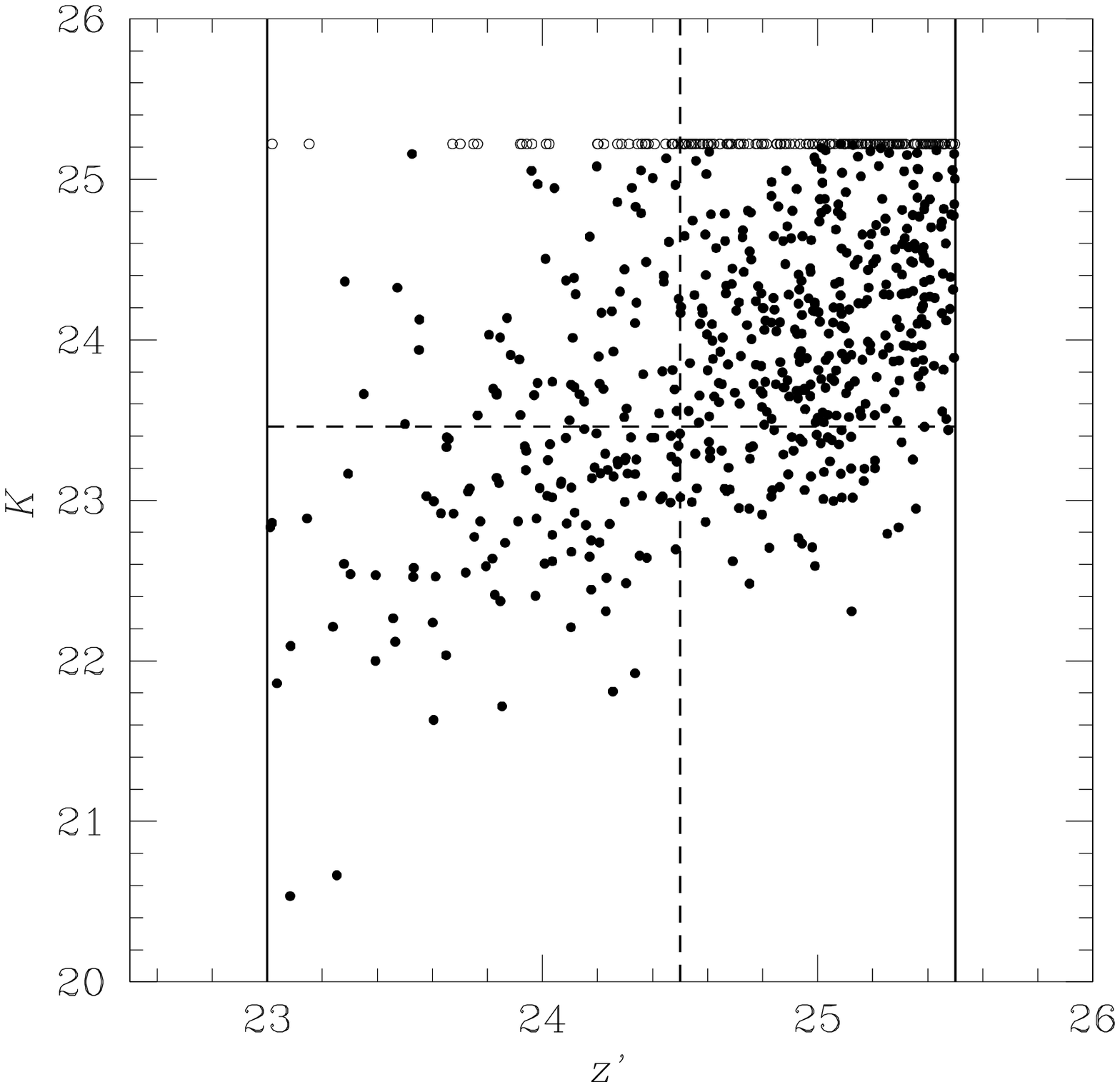}
  \end{center}
  \caption{%
Distribution of the LBG sample on the $z'$ vs $K$ plane.
When the $K$ magnitude of an LBG is fainter
than the $1\sigma$ magnitude,
the $1\sigma$ magnitude is assigned (open circles).
The solid lines indicate the magnitude limit to the sample.
The dashed lines indicate the boundary of four subsamples.
\label{fig:zk}}
\end{figure}

\begin{figure}[t]
  \begin{center}
    \includegraphics[scale=0.4]{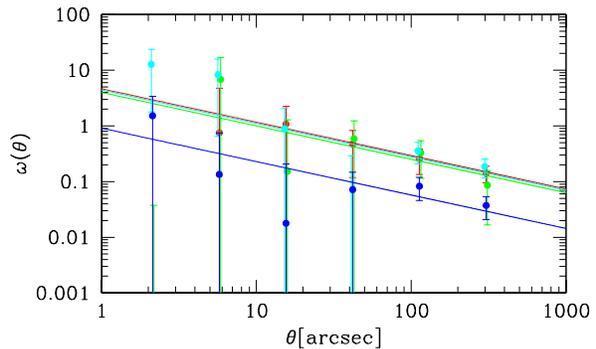}
  \end{center}
  \caption{%
Angular correlation function, $\omega(\theta)$,
for four subsamples selected by $z'$ and $K$ magnitudes.
Colors of the symboles for subsamples are designated as follows:
(red) $23.0 < z' \le 24.5$ and $K \le 23.46$,
(green) $24.5 < z' \le 25.5$ and $K \le 23.46$,
(cyan) $23.0 < z' \le 24.5$ and $K > 23.46$,
and (blue) $24.5 < z' \le 25.5$ and $K > 23.46$.
\label{fig:acf_mag_zk}}
\end{figure}

\begin{figure}[h]
  \begin{center}
    \includegraphics[scale=0.4]{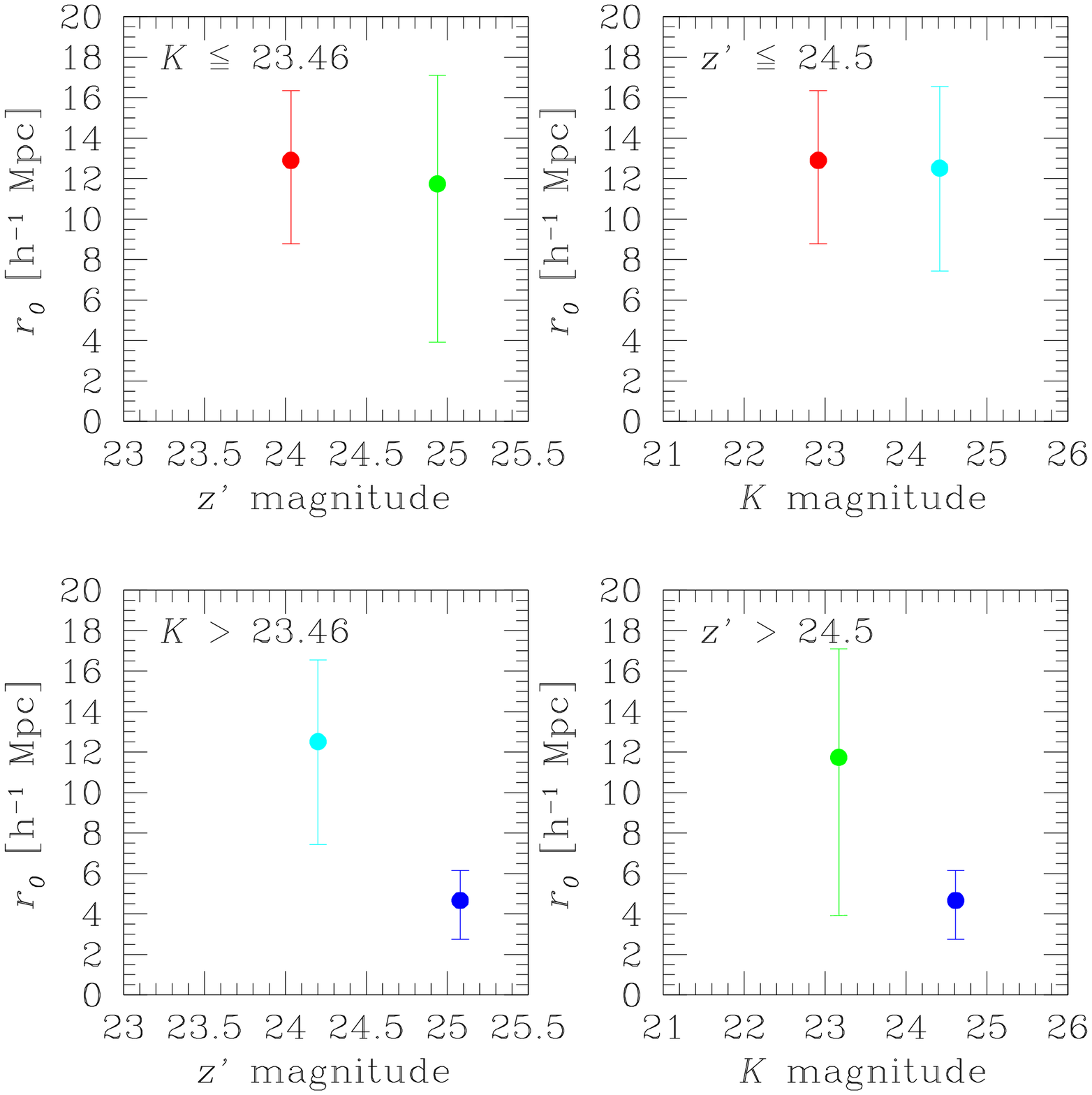}
  \end{center}
  \caption{%
Spatial correlation length, $r_0$, in dependence on $z'$ and $K$
magnitudes.
Colors of the symbols are the same as in \reffig{acf_mag_zk}.
\label{fig:r0_mag_zk}}
\end{figure}

To investigate in what manner the star formation activity of LBGs
relates to the mass of dark halos in detail,
we explore the dependence of the clustering strength
on a combination of $z'$ and $K$ magnitudes,
i.e., the behavior of the clustering strength on the $z'$ vs $K$  plane.
The LBG sample is divided into four subsamples on that plane
($23.0 < z' \le 24.5$ and $K \le 23.46$,
$24.5 < z' \le 24.5$ and $K \le 23.46$,
$23.0 < z' \le 24.5$ and $K > 23.46$,
and $24.5 < z' \le 24.5$ and $K > 23.46$;
\reffig{zk}, \reftbl{clustering}).
The magnitude of $K=23.46$ corresponds to $S/N=5$ in the $K$ band
image.
The ACFs for all the subsamples are presented in \reffig{acf_mag_zk}.
In this figure, we see that only one subsample which is faint
in both $z'$ and $K$ has a smaller correlation amplitude,
whereas the other subsamples which are bright either in $z'$ or $K$
have very similar large correlation amplitudes.
The correlation lengths derived from the amplitudes are plotted
as functions of $z'$ and $K$ magnitudes in \reffig{r0_mag_zk}.
The magnitude of each point in \reffig{r0_mag_zk} is the median
magnitude of the corresponding subsample.
The correlation length of galaxies bright in either magnitude
is large irrespective of the other magnitude.
On the other hand, galaxies faint in one of the two magnitudes
have correlation lengths decreasing with the other magnitude.
The same trends are seen when we further divide the faintest subsample
into two by each of the magnitudes,
although the errors are large due to poor statistics.
The mass of dark halos for these subsamples ranges from
$2\times 10^{13} M_\odot$ for the subsamples with large
correlation amplitudes
to $1\times 10^{12} M_\odot$ for the subsample with small one.

The overall clustering behavior against magnitudes and a color
on the $z'$ versus $K$ plane
suggests that the clustering strength is primarily related to
the combination of $K$ magnitude and $z'$ magnitude
instead of the combination of $K$ magnitude and $z'-K$ color
or $z'-K$ color alone.
For reference to results obtained from
near-infrared selected galaxies,
\citet{quadri2007} examined clustering properties of $K$-selected
galaxies ($K<22.8$)
at $2<z<3.5$
and claimed that the clustering strength
seemed to be independent of $K$ magnitude for their sample.
This does not, however, necessarily conflict with our result,
since the limit of $K$ magnitude of their sample is brighter
than the limit of our brightest subsample.
This is true even for the absolute magnitudes
where the slight difference of redshift range is taken into account.
When we explore the dependence among our brightest subsample
by splitting it into two,
the difference in clustering strength between the two subsamples
is found not to be significant
though there are large errors due to poor statistics.
\citet{ichikawa2007} measured the clustering strength of $K$-selected
galaxies down to $K=25.0$ for a deeper but 28 arcmin$^2$ field
and found that the clustering strength
does increase with $K$-band luminosity at $K>23.0$.
It is indicated that the stellar mass of galaxies may not be a strong
function of the mass of dark halos in the most massive dark halos.
As to the dependence on rest-frame UV luminosity, \citet{quadri2007}
found that the optically brighter subsample clusters less
strongly than the fainter subsample for the $K$-selected galaxies.
However, the subsample that had a larger correlation length
is fainter than $R=25.5$, which is close to the limit of our sample
and we cannot confirm their results from our sample.
They have not explored the dependence for $R<25.5$.
We note here that the samples of \citet{quadri2007}
and \citet{ichikawa2007} are rest-frame optical selected,
while ours is rest-frame UV selected,
and the differences in results may be partly caused by the
difference in the way of selection.

One implication of the result in terms of the stellar mass assembly
of galaxies is that massive halos can host any galaxies
from ones with small stellar masses
to ones which have accumulated large stellar masses,
while in less massive halos,
only galaxies with small stellar masses reside.
There appears to be an upper limit to the stellar mass accumulated
which is determined by the mass of hosting dark halos.
Moreover, galaxies in massive halos can have various star formation
rates whatever stellar masses they have at $z\sim 3$.
On the other hand, galaxies in less massive halos accumulated only
small stellar masses and have lower star formation rates at $z\sim 3$.
\citet{ichikawa2007} suggest a similar tendency for galaxies
with small stellar masses in their $K$-selected galaxy sample
that galaxies red in rest-frame UV color
(galaxies with passive star formation)
in low-mass samples tend to belong to less massive halos,
although their result was derived based on a small field
and contains large errors.
It is suggested that the mass of dark halos governs the
current star formation activity as well as the past star formation
history at $z\sim 3$.

%---------------------------------------------------------------------
%---------------------------------------------------------------------
\section{SUMMARY AND CONCLUSIONS} \label{sec:sum}

We have investigated clustering properties of Lyman-break galaxies
(LBGs) at $z\sim 3$ based on deep multi-waveband imaging data
from optical to near-infrared wavelengths
in the Subaru/\textit{XMM-Newton} Deep Field.
The LBGs are selected by \uv and \vz\ colors in one contiguous area
of 561 arcmin$^2$ down to $z'=25.5$.
The number of LBG candidates detected is 962 in total, among which
$708$ are found in the region observed in the $J$ and $K$ bands.
We use Monte Carlo simulations to estimate the redshift distribution
function and the fraction of contamination by interlopers of the
LBG samples.
The fraction is found to be at most 6\% at any magnitude.

We explored the dependence of the clustering strength on rest-frame
UV and optical magnitudes, which can be indicators of star formation
rate and stellar mass, respectively.
The correlation length is found to be a strong function of
both UV and optical magnitudes with brighter galaxies being
more clustered than faint ones in both cases.
It is found that the correlation length also increases with
$z'-K$ color, which is correlated with $K$ magnitude in our LBG sample.
These results imply that galaxies with larger star formation rates
and larger stellar masses are hosted by more massive dark halos.

Furthermore, the correlation length is interestingly dependent on
a combination of UV and optical magnitudes in the sense that
galaxies bright in optical magnitude have large correlation lengths
irrespective of UV magnitude,
while galaxies faint in optical magnitude have correlation lengths
decreasing with UV magnitude.
One implication of this result in terms of the stellar mass assembly
of galaxies is that galaxies which have accumulated
large stellar masses
always belong to massive halos in which they can have
various star formation
rates, while galaxies with small stellar masses reside in less massive
halos only if they have low star formation rates.
To put it another way, massive halos can host any galaxies
from ones with small stellar masses
to ones with large stellar masses,
while in less massive halos,
only galaxies with small stellar masses reside.
There appears to be an upper limit to the stellar mass accumulated
which is determined by the mass of hosting dark halos.
Moreover, galaxies in massive halos can have various star formation
rates whatever stellar masses they have at $z\sim 3$.
On the other hand, galaxies in less massive halos accumulated only
small stellar masses and have lower star formation rates at $z\sim 3$.
It is suggested that the mass of dark halos governs the
current star formation acitivity as well as the past star formation
history at $z\sim 3$.

\begin{deluxetable}{cccc}[p]
  \tablecolumns{4}
  \tablewidth{0pt}
  \tablecaption{Clustering Measurements \label{tbl:clustering}}
  \tablehead{
    \colhead{\hspace{0.3cm} Sample\hspace{0.3cm} }
    & \colhead{\hspace{0.5cm} $N$\tablenotemark{\ast}\hspace{0.5cm} }
    & \colhead{\hspace{0.3cm} $A_\omega$\hspace{0.3cm} }
    & \colhead{\hspace{0.3cm} $r_0$ ($h^{-1}$Mpc)\hspace{0.3cm} }
  }
  \startdata
$23.0<z'\le 25.5$  & 962 & $1.17^{+0.29}_{-0.29}$ & \phn$5.5^{+0.8}_{-0.9}$ \\
\\
$23.0<z'\le 24.5$ & 288 & $3.44^{+1.00}_{-1.00}$ & $10.7^{+1.9}_{-2.1}$ \\
$24.0<z'\le 25.0$ & 467 & $1.46^{+0.56}_{-0.56}$ & \phn$6.3^{+1.4}_{-1.6}$ \\
$24.5<z'\le 25.5$ & 674 & $0.76^{+0.36}_{-0.35}$ & \phn$4.2^{+1.1}_{-1.3}$ \\
\\
\phantom{$22.96<$}$K\le 23.46$ & 171 & $3.85^{+1.28}_{-1.28}$ & $11.5^{+2.3}_{-2.6}$ \\
$22.96<K\le 23.96$             & 225 & $1.80^{+0.92}_{-0.92}$ & \phn$7.1^{+2.1}_{-2.6}$ \\
$23.46<K$\phantom{$\le 23.96$} & 537 & $0.96^{+0.39}_{-0.39}$ & \phn$4.8^{+1.1}_{-1.3}$ \\
\\
\zk$\le 0.8$ & 365 & $1.14^{+0.52}_{-0.52}$ & \phn$5.4^{+1.4}_{-1.7}$ \\
\zk$>0.8$    & 343 & $2.50^{+0.78}_{-0.79}$ & \phn$8.8^{+1.6}_{-1.9}$ \\
\\
$23.0<z'\le 24.5$,\ \ \ $K\le 23.46$ & 105     & $4.63^{+2.14}_{-2.13}$ & $12.9^{+3.5}_{-4.1}$ \\
$24.5<z'\le 25.5$,\ \ \ $K\le 23.46$ & \phn 66 & $3.98^{+3.29}_{-3.29}$ & $11.7^{+5.4}_{-7.8}$ \\
$23.0<z'\le 24.5$,\ \ \ $23.46<K$    & \phn 96 & $4.41^{+2.49}_{-2.49}$ & $12.5^{+4.0}_{-5.1}$ \\
$24.5<z'\le 25.5$,\ \ \ $23.46<K$    & 441     & $0.91^{+0.51}_{-0.52}$ & \phn$4.7^{+1.5}_{-1.9}$ \\
  \enddata 
  \tablenotetext{\ast\ }{Number of LBGs contained in the sample.}
\end{deluxetable}

%---------------------------------------------------------------------
%---------------------------------------------------------------------
\acknowledgments
We would like to thank the referee for valuable comments and
suggestions.
We are deeply grateful to the Subaru Telescope staff for their
invaluable support in the observations.
M.Y. acknowledges support from the Japan Society for the Promotion of
Science (JSPS) through JSPS Research Fellowship for Young Scientists.

%---------------------------------------------------------------------
%---------------------------------------------------------------------

\end{document}